\documentclass[conference]{IEEEtran}
\pdfoutput=1

\ifCLASSINFOpdf
\else
\fi
\usepackage{graphicx}
\usepackage{url}
\usepackage{amsmath}
\usepackage{comment}
\usepackage{subfigure}
\usepackage{comment}
\usepackage{gensymb}
\usepackage{setspace}
\usepackage{color}
\usepackage{balance}
\usepackage{cite}

\begin{document}
\title{\vspace{-1mm}Near-field 6G Networks: Why Mobile Terahertz Communications MUST Operate in the Near Field\vspace{-3mm}}

\author{Vitaly Petrov,\,\,\,\,\,\,Josep Miquel Jornet\\
Northeastern University, Boston, MA, USA\vspace{-40mm}
\and
Arjun Singh\\
SUNY Polytechnic Institute, Utica, NY, USA\vspace{-40mm}}

\maketitle
\begin{abstract}
Near-field mobile terahertz (THz) communications is one of the candidate enablers for high-rate wireless data exchange in sixth-generation (6G) networks. However, operating in the THz near field brings both attractive opportunities and severe challenges. Hence, it becomes of interest to explore if it is possible to design a realistic mobile THz communication system \emph{without} working in the THz near field. To answer this question, a mathematical framework is presented modeling a mobile THz link that works exclusively in the far field. The study leads to an interesting theoretical conclusion: while the actual frequency is of (almost) no interest, such a system must operate over a limited bandwidth not exceeding a certain threshold. It is then numerically shown that operating only in the far field imposes stringent limitations on mobile THz communications, thus making them less attractive to prospective high-rate services. In contrast, it is shown that a stationary THz link can still be broadband even when staying exclusively in the THz far field. Hence, broadband mobile THz communications MUST be near-field, while broadband stationary THz links do not have to.
\end{abstract}
\IEEEpeerreviewmaketitle

\vspace{-1.35mm}
\section{Introduction}
\vspace{-1.2mm}
\label{sec:intro}
In recent years, terahertz (THz, $0.3$\,THz--$10$\,THz) and sub-THz ($100$\,GHz--$300$\,GHz) communications have been almost unanimously accepted as being a key enabling technology for future wireless standards~\cite{rappaport2019wireless}. Large bands available at (sub-)THz frequencies can be utilized to design very high data rate and low latency networks. With the THz technology gap closing, THz radios have been utilized in demonstrating wideband~\cite{sen2020teranova}, long range~\cite{sen2023multi}, and high data rate links~\cite{kuerner_thor}.

With the feasibility of stationary (sub-)THz communications thus proven, the research focus shifts to exploring mobile (sub-)THz links as a candidate solution for sixth-generation (6G) networks. Such broadband mobile THz links may be used for bandwidth-hungry extended reality (XR) services~\cite{xr_thz}, joint communications and sensing~\cite{sarieddeen2020next}, airborne networks~\cite{chong_uav}, and CubeSat mega constellations~\cite{alqaraghuli2021performance}, among others.

A novel complication that arises at THz is from the fact that the very high gain aperture antennas, arrays, or reconfigurable intelligent surfaces (RISs) required to overcome the huge path losses at THz frequencies also exhibit \emph{a large near-field zone}~\cite{balanis2016antenna}. On the one hand, operating in the near field leads to novel opportunities, including the possibility to exploit novel wavefronts, such as beamfocusing~\cite{nf_1960_2}, Bessel beams~\cite{durnin1988comparison}, and Airy beams~\cite{siviloglou2007observation}, among others. Here, novel attractive designs become possible, not feasible within far-field systems.

On the other hand, operation in the THz near field also leads to a number of challenges~\cite{near_field_6g_magazine}. First, the propagation of the wave has to be modeled with care, as the phase of the wavefront cannot be ignored as with a far-field planar wave approximation. Second, some fundamental assumptions from the far field do not always hold, such as the canonical free space path loss (FSPL) equation as per the Friis transmission formula~\cite{balanis2016antenna}. In addition, canonical beamforming heavily used in 4G/5G microwave and 5G mmWave communications is much less efficient in the near field~\cite{near_field_6g_magazine,balanis2016antenna,singh2023wavefront}. Beamforming ``focuses" the beam at infinity, hence a ``pencil-sharp'' far-field beam becomes wider in the near field. The mobility of nodes between the near-field and far-field zones imposes another complication~\cite{chong_magazine_nf_ff}. All these complicate the design of an efficient near-field THz communication system~\cite{petrov4mobile}.

This combination of challenges leads to the following questions: \emph{Is it possible to design an efficient THz communication system without dealing with near-field propagation?}

\emph{In this paper,} we answer this question as we study the need to operate in the near field for both stationary and non-stationary (mobile) THz communication systems. We prove by contradiction -- by modeling an idealistic exclusively far-field THz communication system. Our main theoretical result is that such a system appears to be limited not by the central frequency or wavelength (used in the canonical Fraunhofer near field distance~\cite{emil_2021_asilomar}), but rather by its maximum bandwidth.

Our main engineering conclusion is that the bandwidth limit for stationary THz links is very high, while the value for a mobile THz link may be even lower than the existing 5G mmWave bands. We thus show that an exclusively far-field non-stationary THz link cannot provide acceptable performance metrics, so mobile (sub-)THz communication systems will have to (at least partially) operate in the near field.

There have been several recent useful theoretical studies on near-field 6G communications (\cite{emil_2021_asilomar,nf_pisa,emil_power,nf_localization,chong_technical}, among others), following the fundamental works from 1960s~\cite{nf_1960_1,nf_1960_2}. However, they do not directly connect (i)~\emph{mobility} with (ii)~the \emph{operation in the near-field zone}, which, as we learn, is an important connection to make for broadband THz communications.

\vspace{-1.5mm}
\section{System Model}
\begin{figure}[!t]
\centering
 \includegraphics[width=0.9\columnwidth]{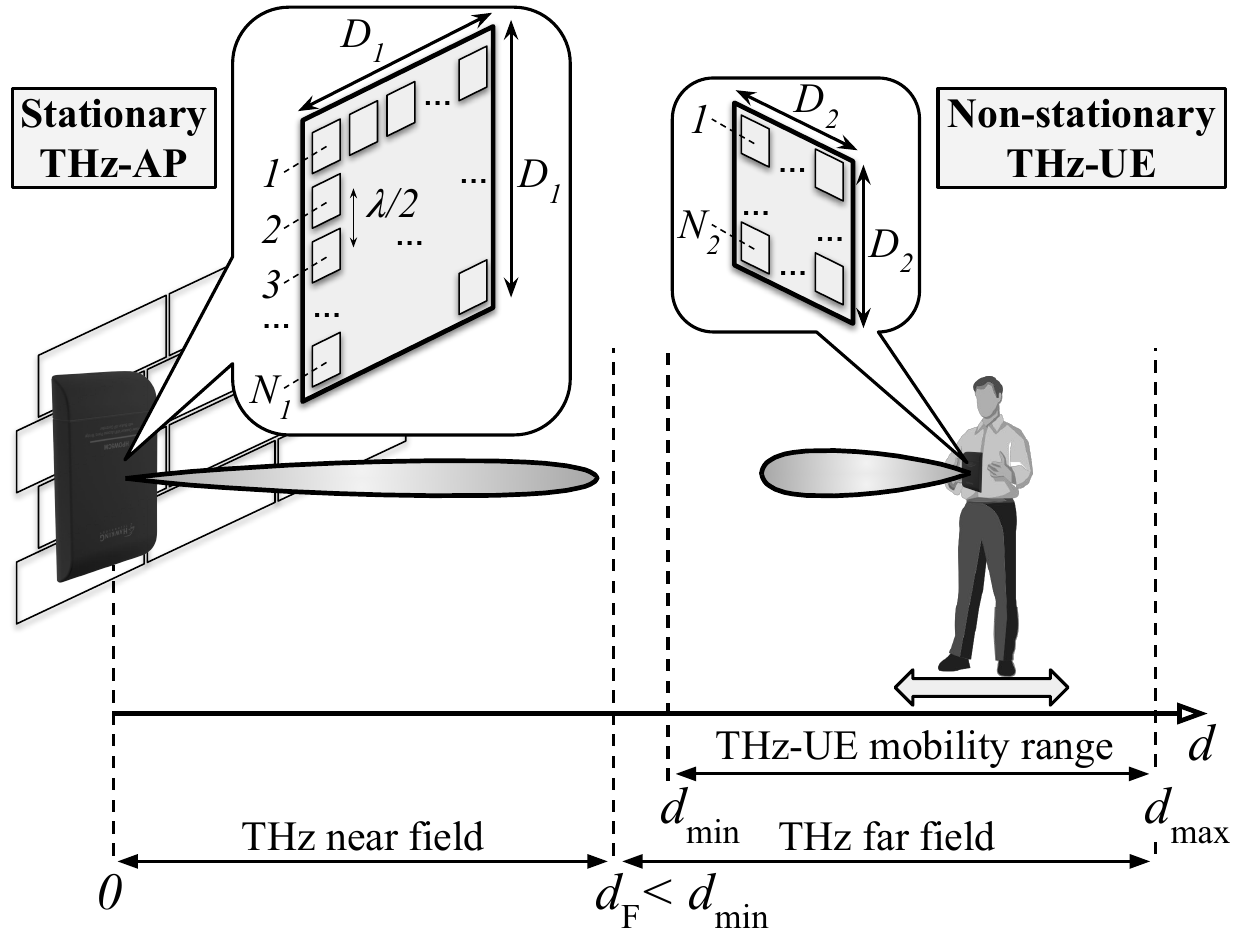}
 \vspace{-3mm}
 \caption{Non-stationary THz link working exclusively in the THz far field.}
 \label{fig:setup}
  \vspace{-7mm}
\end{figure}

\vspace{-1mm}
\label{sec:system_model}
Our system is illustrated in Fig.~\ref{fig:setup}. We model a point-to-point non-stationary THz communication link between a stationary THz Access Point (THz-AP) and a mobile THz User Equipment (THz-UE). Both sides are equipped with planar THz antenna arrays with $\lambda/2$ spacing in-between the elements, where $\lambda$ is the wavelength corresponding to the central frequency $f$, so $\lambda = c/f$. The THz-AP has an array of $N_{1} \times N_{1}$ elements and physical dimensions of $D_{1} \times D_{1}$\,m$^2$, while the corresponding array at the THz-UE side is of $N_{2} \times N_{2}$ elements and $D_{2} \times D_{2}$\,m$^2$, respectively.

The THz-AP and the THz-UE are separated by the distance $d$ varying from a certain minimal distance, $d_{\text{min}}$\footnote{In practical deployments, this value may be determined by the safety limits or the height of the THz-AP mounted on a lamppost or a wall.}, to a maximum communication range, $d_{\text{max}}$, determining the coverage of the THz-AP. To simplify the derivations presented below, we present the case with only the mobility of the THz-UE along the line $[d_{\text{min}}, d_{\text{max}}]$, so the antenna arrays are assumed to always be broadside to each other. The analysis can be extended by also considering the mutual rotations. Importantly, the current configuration provides the maximum gain from an array and thus satisfies the most stringent requirement for the near-field zone discussed in the next section. \emph{Therefore, modeling just the variable separation distance for mobile THz links is sufficient to prove our main points.}

The transmit (Tx) power $P_{\text{Tx}}$ is equally spread over the signal bandwidth $B$. The receiver (Rx) side also has a noise figure $N_{\text{F}}$ to account for additional noise above the thermal noise. The resulting communication model in Section~\ref{sec:analysis} is for both uplink and downlink: in downlink, $P_{\text{Tx}}$ stands for the THz-AP Tx power and $N_{\text{F}}$ -- for the THz-UE noise figure; in uplink, $P_{\text{Tx}}$ stands for the THz-UE Tx power and $N_{\text{F}}$ -- for the THz-AP noise figure. We introduce two additional coefficients for our analysis: (i) the mobility coefficient $M = d_{\text{max}} / d_{\text{min}}$; and (ii)~the antenna inequality coefficient $L = D_{1} / D_{2}$.

\vspace{-2mm}
\section{Analysis: Proof by Contradiction}
\label{sec:analysis}
\vspace{-1mm}

In this study, we analyze the limits of a communication system designed to operate exclusively in the far field, and thus, whenever the system requirements are contradicted, this proves the need for near-field THz communications.

\vspace{-1mm}
\subsection{General Case and the Limit for a Stationary THz Link}
\label{sec:general}
The canonical boundary between the near and far field of a radiating aperture is given per the Fraunhofer distance, $d_{\text{F}}$~\cite{balanis2016antenna}. This distance demarcates the region beyond which the plane wave approximation yields a phase discrepancy of $\leq$$\pi/8$ in the planar wavefront as compared to the true wavefront under spherical wave propagation. For all the distances $d \geq d_{\text{F}}$, the far-field zone is applicable, while $d < d_{\text{F}}$ is the near field~\cite{balanis2016antenna}.

For prospective THz communication systems, for the case with a Tx and Rx equipped with a $D_{1} \times D_{1}$ and $D_{2} \times D_{2}$ array respectively, the Fraunhofer distance $d_{\text{F}}$ is given as:
\begin{align}
\vspace{-5mm}
d_{\text{F}} = 4(D_{1} + D_{2})^2 / \lambda.
\label{eq:d_f_arrays}
\vspace{-1mm}
\end{align}
A discussion on how (\ref{eq:d_f_arrays}) is derived is in Appendix~\ref{app:fraunhofer_distance_array}\footnote{Possibly, (\ref{eq:d_f_arrays}) is not presented for the first time ever (i.e., similar cases are discussed in~\cite{near_field_6g_magazine} and~\cite{emil_2021_asilomar}). We just need an accurate equation to proceed and haven't found a reference to cite it in this exact form (for two square arrays).}.

To always stay in the far field, the minimal communication range $d_{\text{min}}$ must then fulfill the following condition:
\begin{align}
\vspace{-1mm}
d_{\text{min}} \geq 4(D_{1} + D_{2})^2 / \lambda,
\label{eq:condition1}
\vspace{-1mm}
\end{align}
which we refer to as \emph{Condition~1}. Hence, $(D_{1} + D_{2}) \leq \frac{\sqrt{\lambda d_{\text{min}}}}{2}$.

However, the system must also satisfy \emph{Condition~2}:
\begin{align}
S(d_{\text{max}}) \geq S_{\text{L}},
\label{eq:condition2_orig}
\end{align}
where $S(d_{\text{max}})$ is the signal-to-noise ratio (SNR) at $d = d_{\text{max}}$, while $S_{\text{L}}$ is the threshold SNR (set per system requirements). 

We then recall that the linear antenna gains $G_{1}$ and $G_{2}$ can be computed for square arrays as $G_{1} = 4\pi D^2_{1}/\lambda^{2}$ and $G_{2} = 4 \pi D^2_{2}/\lambda^{2}$, respectively~\cite{balanis2016antenna}. Thus, (\ref{eq:condition2_orig}) is reformulated~as\footnote{We model a line-of-sight (LoS) channel in the THz transparency window, so the molecular absorption is of secondary importance compared to the spreading loss, transmit power, and antenna gains.}:
\begin{align}
&\quad{}\quad{}\frac{P_{\text{Tx}} G_{1} G_{2}}{N_{0}} \Big( \frac{\lambda}{4 \pi d_{\text{max}}} \Big)^2 \geq S_{\text{L}}\quad{}\Rightarrow \nonumber\\
&\Rightarrow\quad{}\frac{P_{\text{Tx}} \frac{4 D^2_{1}}{\lambda^2} \frac{4 D^2_{2}}{\lambda^2} \frac{\lambda^2\pi^2} {16 \pi^2 d^2_{\text{max}}}}{N_{0}} \geq S_{\text{L}}\quad{}\Rightarrow \nonumber\\
&\Rightarrow\quad{}D_{1} D_{2} \geq \lambda d_{\text{max}} \sqrt{ \frac{N_{\text{F}} k T B}{P_{\text{Tx}}} } 10^{\frac{S_{\text{L}, \text{dB}}}{20}},
\label{eq:condition2}
\end{align}
where $N_{0}$ = $BN_{\text{F}}kT$, with $N_{\text{F}}$ as the noise factor, $k$ the Boltzmann constant, $T$ the system temperature in Kelvin. $B$ stands for the bandwidth of the transmitted signal in Hz, while $S_{\text{L},\text{dB}}$ is the SNR threshold, $S_{\text{L}}$, in dB scale.

\begin{figure*}[!t]
\vspace{-2mm}
\setcounter{equation}{15}
\begin{align}
P_{\text{Tx}, \text{dBm}} = 54.08 + S_{\text{L}, \text{dB}} + N_{\text{F}, \text{dB}} + 10\log_{10}\big(kT\big) + 10\log_{10}\big( B^{\text{(max)}}_{\text{Mobile}} \big) + 20\log_{10}\big( M \big) + 20\log_{10}\bigg( \frac{(L+1)^2}{4L} \bigg)
\label{eq:p_Tx_final_mobile}
\end{align}
\normalsize
\setcounter{equation}{4}
\vspace{-8mm}
\end{figure*}

Thus, a reliable mobile THz system as in Section~\ref{sec:system_model} must then satisfy both \emph{Condition~1} and \emph{Condition~2}:
\begin{align}
\begin{cases}
D_{1} + D_{2} &\leq \frac{\sqrt{\lambda d_{\text{min}}}}{2}\\
\,\,\,D_{1} D_{2} &\geq \lambda d_{\text{max}} \sqrt{ \frac{N_{\text{F}} k T B}{P_{\text{Tx}}} } 10^{\frac{S_{\text{L}, \text{dB}}}{20}}.
\end{cases}
\label{eq:cases_orig}
\end{align}

From~\eqref{eq:cases_orig}, we can solve for $D_{1}$ by setting $D_{2} = \frac{\sqrt{\lambda d_{\text{min}}}}{2} - D_{1}$:
\begin{align}
D_{1} \big( \sqrt{\lambda d_{\text{min}}} - 2 D_{1} \big) \geq 2 \lambda d_{\text{max}} \sqrt{ Z } 10^{\frac{S_{\text{L}, \text{dB}}}{20}},
\label{eq:cond2_pre_square}
\end{align}
where $Z$ is set to $Z = \frac{N_{\text{F}} k T B}{P_{\text{Tx}}}$ to simplify the notation.

Then, \eqref{eq:cond2_pre_square} converts into a monic quadratic inequation for~$D_{1}$:
 \begin{align}
D^2_{1} - D_{1}\frac{\sqrt{\lambda d_{\text{min}}}}{2} + \lambda d_{\text{max}} \sqrt{ Z } 10^{\frac{S_{\text{L}, \text{dB}}}{20}} \leq 0,
 \label{eq:cond2_square}
 \end{align}
that can be rewritten as $D^2_{1} + p D_{1} + q \leq 0$, where $p = - \frac{\sqrt{\lambda d_{\text{min}}}}{2}$ and $q = \lambda d_{\text{max}} \sqrt{ Z } 10^{\frac{S_{\text{L}, \text{dB}}}{20}}$.

The presented quadratic inequation has up to two roots, $x_{1}$ and $x_{2}$, where $x_{1} \leq x_{2}$. Then, as the expression must be less or equal to $0$, the legitimate values for $D_{1}$ that satisfy this inequation are $D_{1} \in [x_{1}, x_{2}]$. Since $D_{1}$ specifies the size of the antenna array, $D_{1} > 0$ must be always enforced. Therefore, the sign of $x_{1}$ (if it exists) becomes of interest to determine the valid range of $D_{1}$. For this purpose, we utilize Vieta's formulas connecting the roots of the polynomial to its coefficient. For the quadratic case, if at least one root exists, we get:
\begin{align}
\begin{cases}
x_{1} + x_{2} = - p = \frac{\sqrt{\lambda d_{\text{min}}}}{2}\\
x_{1} x_{2} = q = \lambda d_{\text{max}} \sqrt{ Z } 10^{\frac{S_{\text{L}, \text{dB}}}{20}},
\end{cases}
\label{eq:vieta}
\end{align}
where $p$ and $q$ are the polynomial coefficients defined above.

Analyzing the right sides of (\ref{eq:vieta}), we conclude that both $x_{1}$ and $x_{2}$ (if they exist) must be non-negative\footnote{Here, we notice that $q$ is always positive, as there are no negative (or even zero-value) multipliers in it. Hence, if at least one of the roots $x_{1}$, $x_{2}$ exist, they are either both positive or both negative. However, they cannot be both negative as $x_{1} + x_{2} = - p$ is always non-negative, as $\sqrt{\lambda d_{\text{min}}}/4$ does not have negative multipliers either (here, we say \emph{non-negative} instead of \emph{positive}, as, strictly speaking, $d_{\text{min}}$ may be~$0$). So, both $x_{1}$ and $x_{2}$ are non-negative. Hence, as $x_{1} \geq 0$, $D_{1} > 0$ is always enforces if $D_{1} \in [x_{1}, x_{2}]$.}. Therefore, $D_{1} \in [x_{1}, x_{2}]$ is sufficient to determine $D_{1}$ that satisfies (\ref{eq:cond2_square}).

Here, let us recall that for the quadratic polynomial, the existence of at least one possible root is determined by the sign of its discriminant $\left(p/2\right)^2 - q$. \emph{Hence, there is always at least one value for $D_{1}$ that satisfies both Condition 1 and Condition 2 (so that the modeled mobile THz system is both far-field and reliable) \underline{if and only if} $\left(p/2\right)^2 - q$ is non-negative:}
\begin{align}
&\exists D_{1}: \Big( D^2_{1} - D_{1}\frac{\sqrt{\lambda d_{\text{min}}}}{2} + \lambda d_{\text{max}} \sqrt{ Z } 10^{\frac{S_{\text{L}, \text{dB}}}{20}} \Big) \leq 0\quad{}\Leftrightarrow\nonumber\\
&\Leftrightarrow\quad{}\Big( \frac{\sqrt{\lambda d_{\text{min}}}}{4} \Big)^2 - \lambda d_{\text{max}} \sqrt{ Z } 10^{\frac{S_{\text{L}, \text{dB}}}{20}} \geq 0
\label{eq:discriminant}
\end{align}

From (\ref{eq:discriminant}), the following restriction for the signal bandwidth, $B$, is formulated:
\begin{align}
B \leq \left(P_{\text{Tx}} d^2_{\text{min}}\right) / \left(256 d^2_{\text{max}} 10^{\frac{S_{\text{L}, \text{dB}}}{10}} N_{\text{F}} k T\right).
\label{eq:bw_general}
\end{align}

The widest bandwidth constrained by \eqref{eq:bw_general} can be achieved when \eqref{eq:discriminant} has just one root, so $D_{1} = x_{1} = x_{2}$. We now recall \eqref{eq:vieta} determining that $x_{1} + x_{2} = \frac{\sqrt{\lambda d_{\text{min}}}}{2}$. Hence, the actual value of $D_{1}$ leading to the widest achievable bandwidth in \eqref{eq:bw_general} is:
\begin{align}
D_{1} = x_{1} = x_{2} = \sqrt{\lambda d_{\text{min}}} / 4.
\label{eq:D_1_best}
\end{align}
Then, from (\ref{eq:cases_orig}), the max value of $D_{2}$ is $D_{2} = \sqrt{\lambda d_{\text{min}}}/4$. Hence, the best solution for the joint \emph{Condition~1} and \emph{Condition~2} requires symmetric Tx and Rx antennas, or $D_{1} = D_{2}$.

Note that the case, where $D_{1} = D_{2}$, is not typical to mobile environments, where, by default, the antenna size at the mobile user is much smaller than the one employed by the access point, $D_{2} \ll D_{1}$. Thus, this setup of  $D_{1} = D_{2}$ is more relevant to a stationary setup (i.e., a THz wireless backhaul link) featured by no mobility: $d_{\text{min}} = d_{\text{max}} = d$ (see Fig.~\ref{fig:setup}). 

This leads to the following equation for the maximum bandwidth of a stationary THz link, $B^{\text{(max)}}_{\text{stationary}}$, equipped with two equal-size antenna arrays on both sides and a single communication range $d$ that does not change:
\begin{align}
B^{\text{(max)}}_{\text{Stationary}} = \frac{1}{ 256 k T } \cdot 10^{\frac{ P_{\text{Tx}, \text{dBm}} - S_{\text{L}, \text{dB}} - N_{\text{F}, \text{dB}} - 30 }{10}}
\label{eq:bw_final_stationary},
\end{align}
where $P_{\text{Tx},\text{dBm}}$ is the transmit power in dBm, $S_{\text{L},\text{dB}}$ is the SNR threshold in dB, and $N_{\text{F}, \text{dB}}$ is the Rx's noise figure in dB.

\emph{This is an important theoretical result highlighting that as long as a \underline{stationary} THz system is designed without exceeding the maximum bandwidth in~\eqref{eq:bw_final_stationary}, it can operate in the far field.}


\subsection{Limit for Mobile THz Communications}
\label{sec:general}


In general, the value of $D_{2}$ is primarily subject to the form factor of the THz-UE and may be fixed to a certain maximum value $D_{2} = D^{\text{(max)}}_{2}$for different categories of devices, i.e., a smartphone, a tablet, or XR glasses~\cite{xr_thz}. When $D_{2}$ is fixed to $D^{\text{(max)}}_{2}$, a similar analysis for \eqref{eq:cases_orig} follows. First, the corresponding value of $D_{1}$ is found as $D_{1} = \frac{\sqrt{\lambda d_{\text{min}}}}{2} - D^{\text{(max)}}_{2}$. Utilizing this then leads to the following limitation for the maximum achievable bandwidth of a mobile THz link, while satisfying both the far-field operation requirement, \emph{Condition~1}, and the reliability constraint, \emph{Condition~2}:
\begin{align}
B^{\text{(Fixed } D_{2}\text{)}}_{\text{Mobile}} \leq \frac{Q^2}{ 4 k T \lambda^2 d^2_{\text{max}}} \cdot 10^{\frac{ P_{\text{Tx}, \text{dBm}} - S_{\text{L}, \text{dB}} - N_{\text{F}, \text{dB}} - 30 }{10}},
\label{eq:bw_final_mobile_fixed_D2}
\end{align}
where $Q =D^{\text{(max)}}_{2} \big( \sqrt{\lambda d_{\text{min}}} - 2 D^{\text{(max)}}_{2} \big)$ simplifies the notation.

The value in (\ref{eq:bw_final_mobile_fixed_D2}) presents a correct limit for a mobile case with the size of the THz-UE antenna limited by a certain value $D^{\text{(max)}}_{2}$. However, analyzing this equation numerically imposes a challenge, as, in contrast to (\ref{eq:bw_final_stationary}), the achievable bandwidth is now subject to as many as \emph{seven individual parameters}: $d_{\text{min}}$ and $d_{\text{max}}$ (distances), $\lambda$ (wavelength), $P_{\text{Tx}, \text{dBm}}$, $S_{\text{L}, \text{dB}}$, and $N_{\text{F}}$ (radio parameters) and also $D^{\text{(max)}}_{2}$ (THz-UE size limit).

To simplify the analysis, a possible solution here is to reformulate the problem; instead of limiting the THz-UE antenna size to a certain value, we set it to be $L$ times smaller than the corresponding size of the antenna at the THz-AP\footnote{Strictly speaking, these two formulations are not $100\%$ equivalent to each other, so the results differ slightly. Still, the numerical difference is not drastic and is well compensated by a notably simpler formulation presented below.}.

Now, by incorporating the \emph{antenna inequality coefficient} $L = D_{1} / D_{2}$ and the \emph{mobility coefficient} $M = \frac{d_{\text{max}}}{d_{\text{min}}}$ into \eqref{eq:cases_orig}, we derive the bandwidth limit for a mobile THz system, $B_{\text{Mobile}}$:
\begin{align}
B_{\text{Mobile}} \leq \frac{L^2}{ 16 k T M^2 (L+1)^4} \cdot 10^{\frac{ P_{\text{Tx}, \text{dBm}} - S_{\text{L}, \text{dB}} - N_{\text{F}, \text{dB}} - 30 }{10}}.
\label{eq:bw_final_mobile_ineq}
\end{align}

We observe a similarity between \eqref{eq:bw_final_stationary} and \eqref{eq:bw_final_mobile_ineq} leading to the following maximum bandwidth of a mobile THz system that satisfies both the SNR and the far-field requirements:
\begin{align}
B^{\text{(max)}}_{\text{Mobile}} = B^{\text{(max)}}_{\text{Stationary}} / \Big( M^2 \frac{(L+1)^4}{16 L^2} \Big).
\label{eq:bw_final_mobile}
\end{align}

\emph{Thus, the bandwidth of a mobile system is always smaller, affected by \underline{two penalties}}. The first penalty is $M^2$ determined by the range in the communication distances, $M = d_{\text{max}} / d_{\text{min}}$. The second penalty comes from the THz-UE antenna being $L$ times smaller than the one at the THz-AP, $L = D_{1} / D_{2}$.

Reversing (\ref{eq:bw_final_mobile}) and (\ref{eq:bw_final_stationary}), the required power $P_{\text{Tx}}$ to deliver a certain target bandwidth $B$ of a mobile THz link is in (\ref{eq:p_Tx_final_mobile}).

\begin{figure*}[t!]
\centering
\vspace{3mm}
 \subfigure[Effect of the transmit power]
 {
  \includegraphics[height=0.235\textwidth]{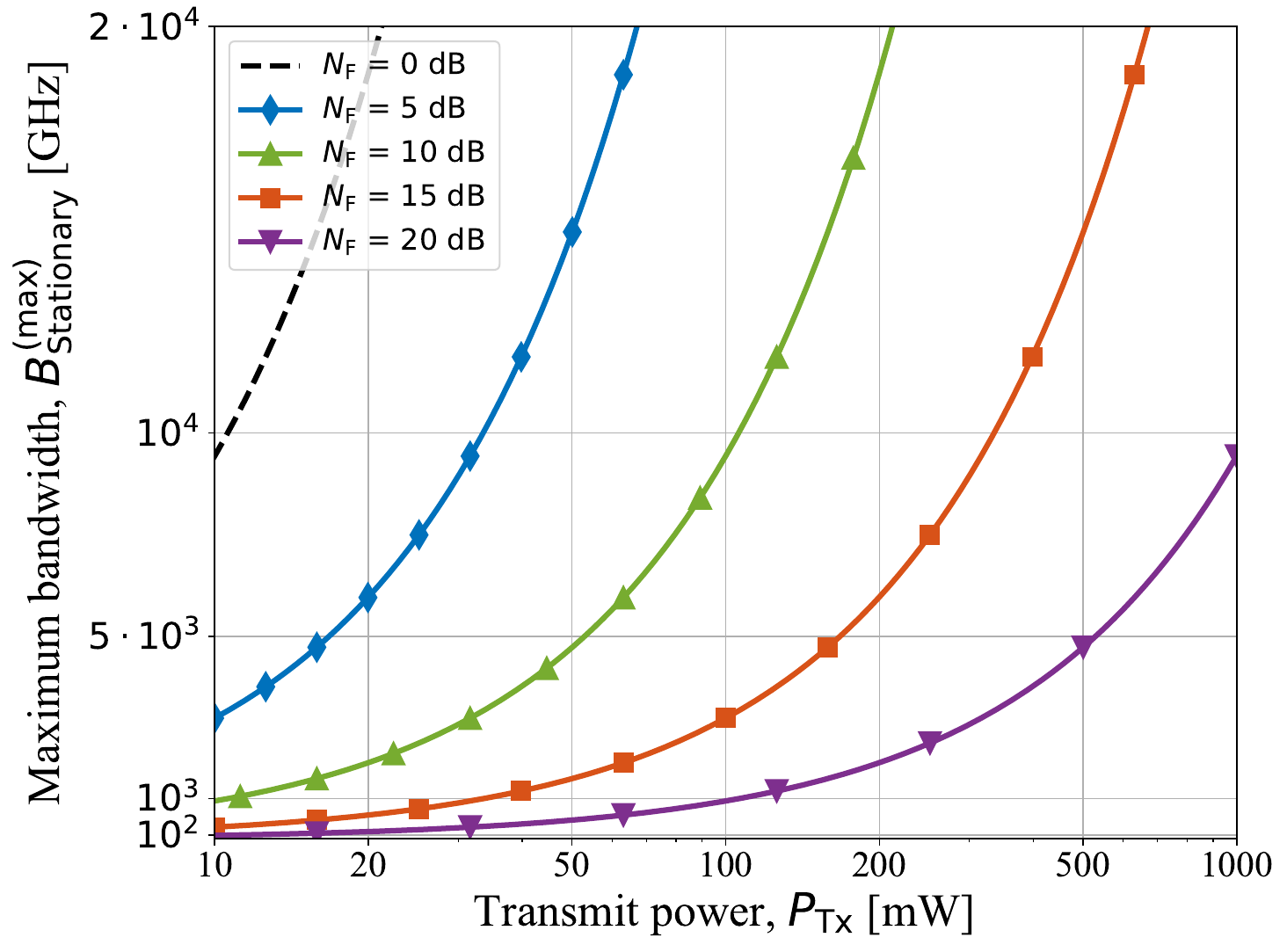}
  \label{fig:plot1}
 }
 \subfigure[Effect of the SNR threshold]
 {
  \includegraphics[height=0.235\textwidth]{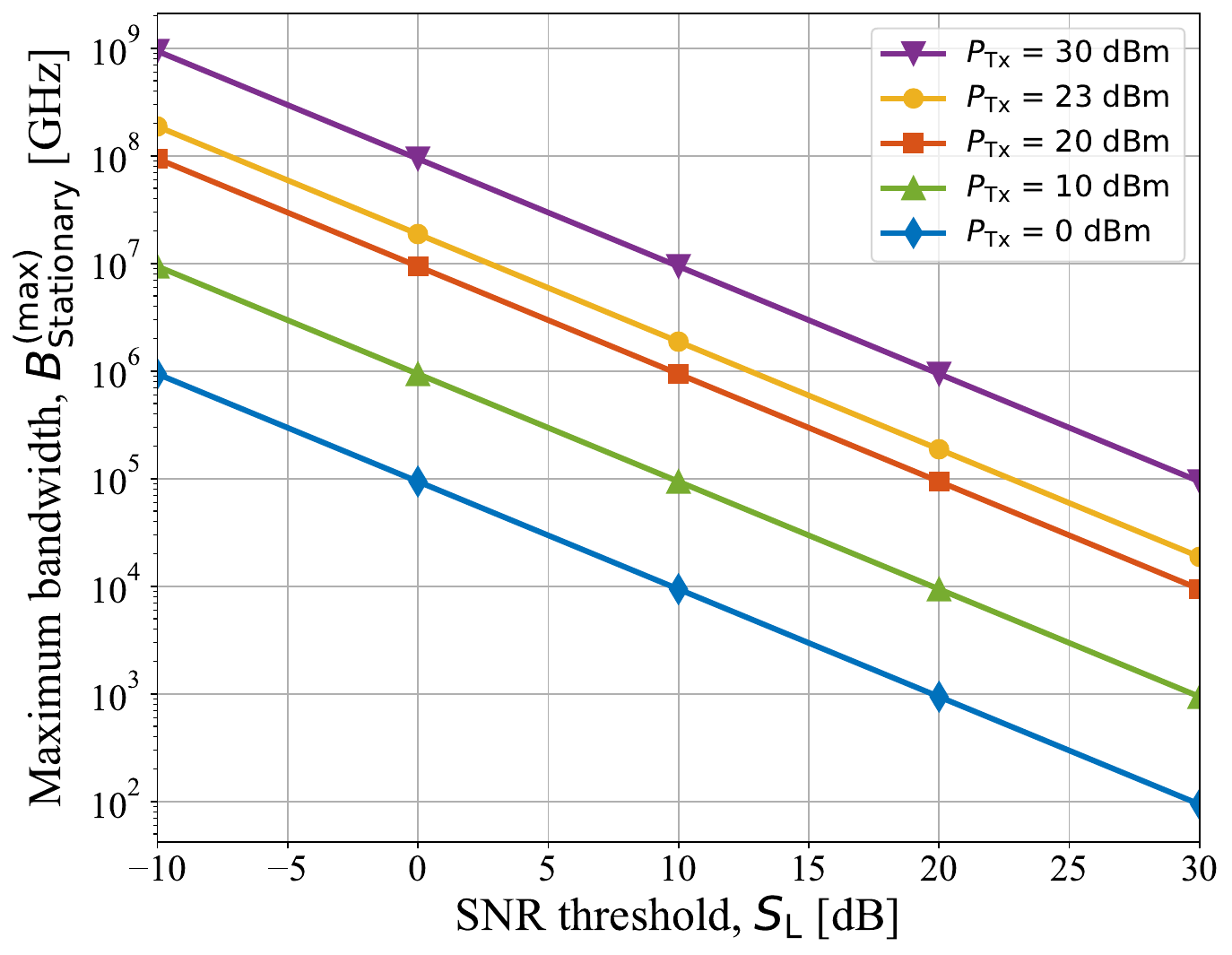}
  \label{fig:plot2}
 }
  \subfigure[Effect of the central frequency]
 {
  \includegraphics[height=0.235\textwidth]{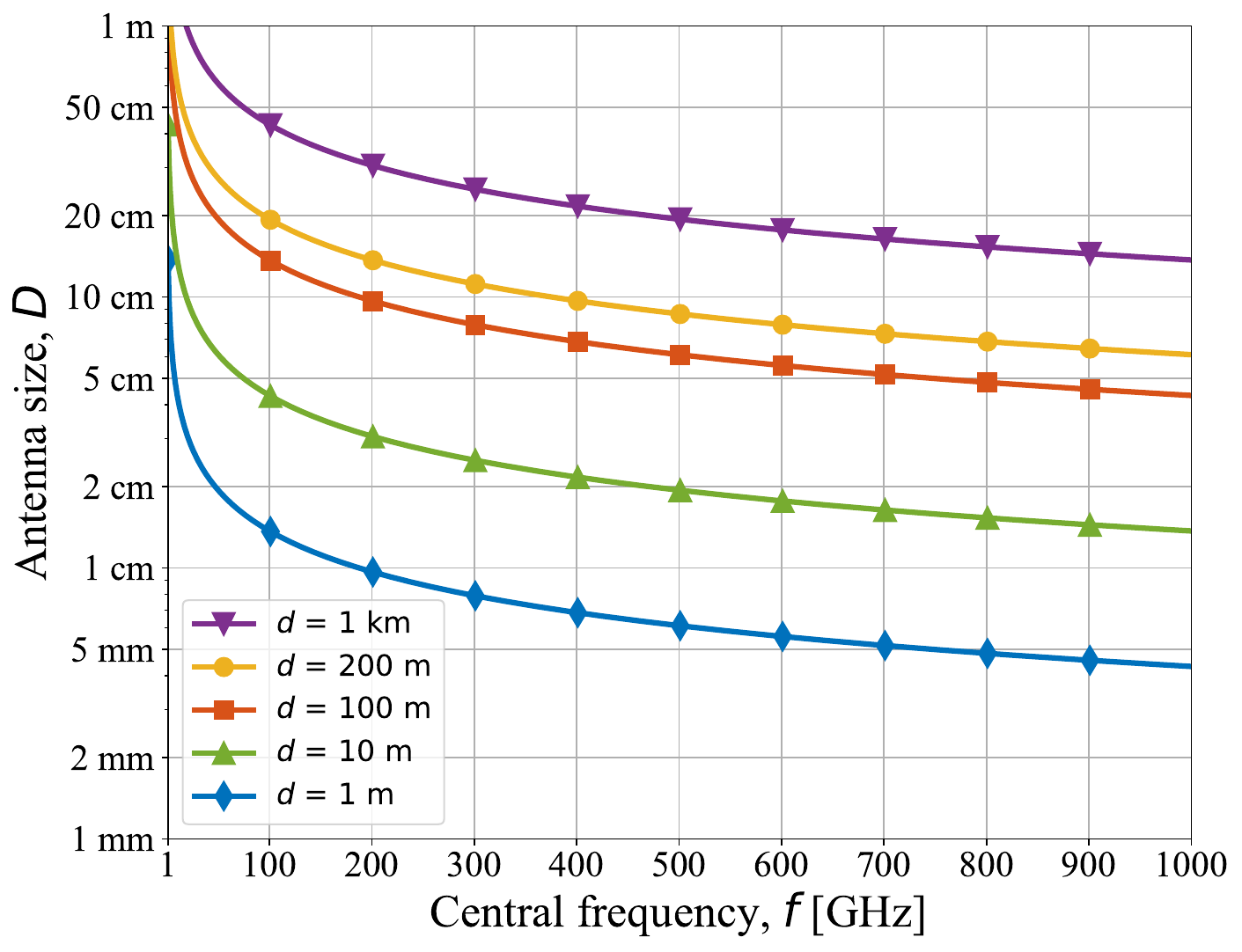}
  \label{fig:plot3}
 }
 \vspace{-2mm}
 \caption{The maximum achievable bandwidth for the far-field \emph{stationary} THz link and the corresponding antenna sizes.}
 \label{fig:stationary}
 \vspace{-3mm}
\end{figure*}

\begin{figure*}[t!]
\centering
 \subfigure[Effect of mobility]
 {
  \includegraphics[height=0.243\textwidth]{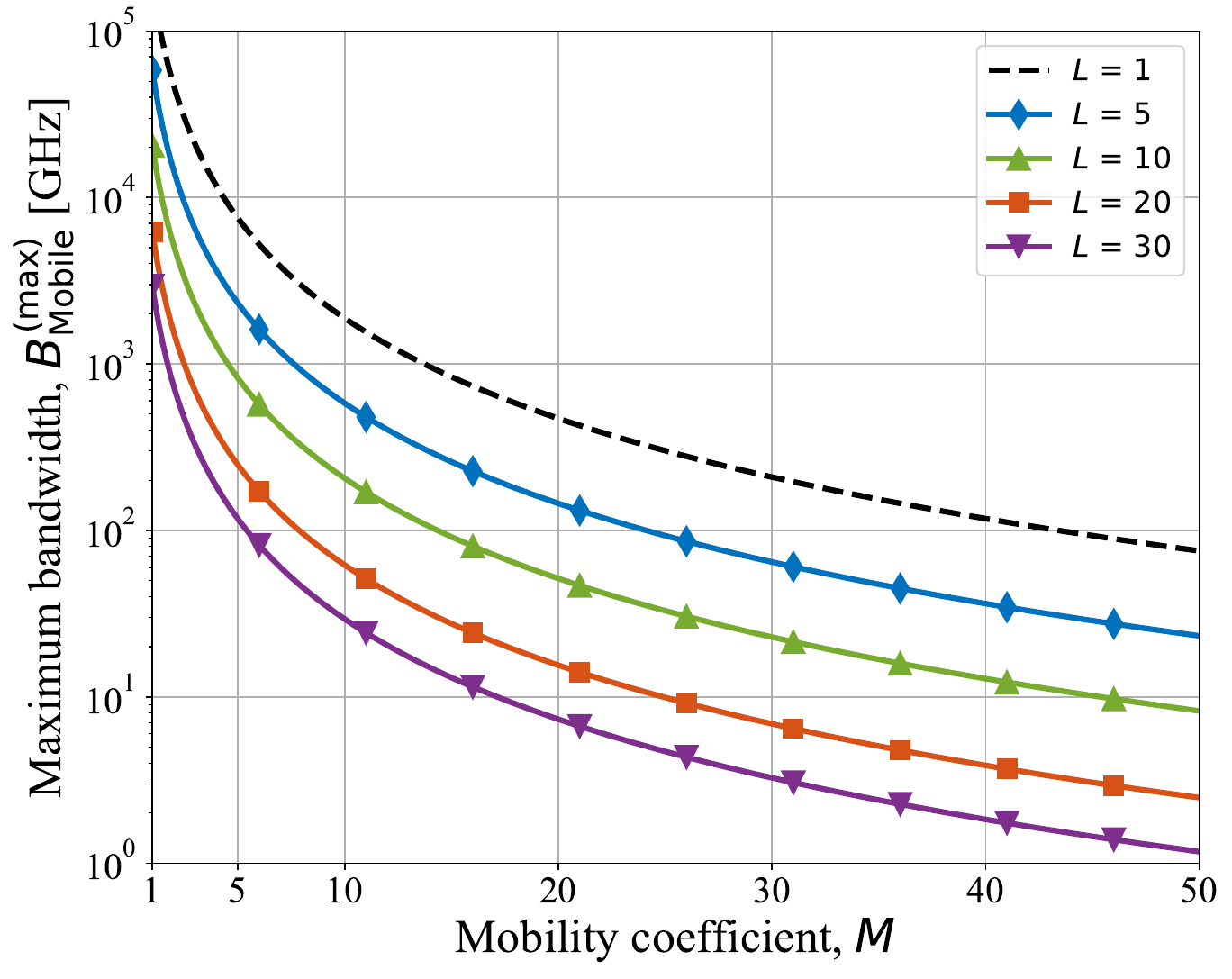}
  \label{fig:plot4}
 }
 \subfigure[Effect of antenna inequality]
 {
  \includegraphics[height=0.243\textwidth]{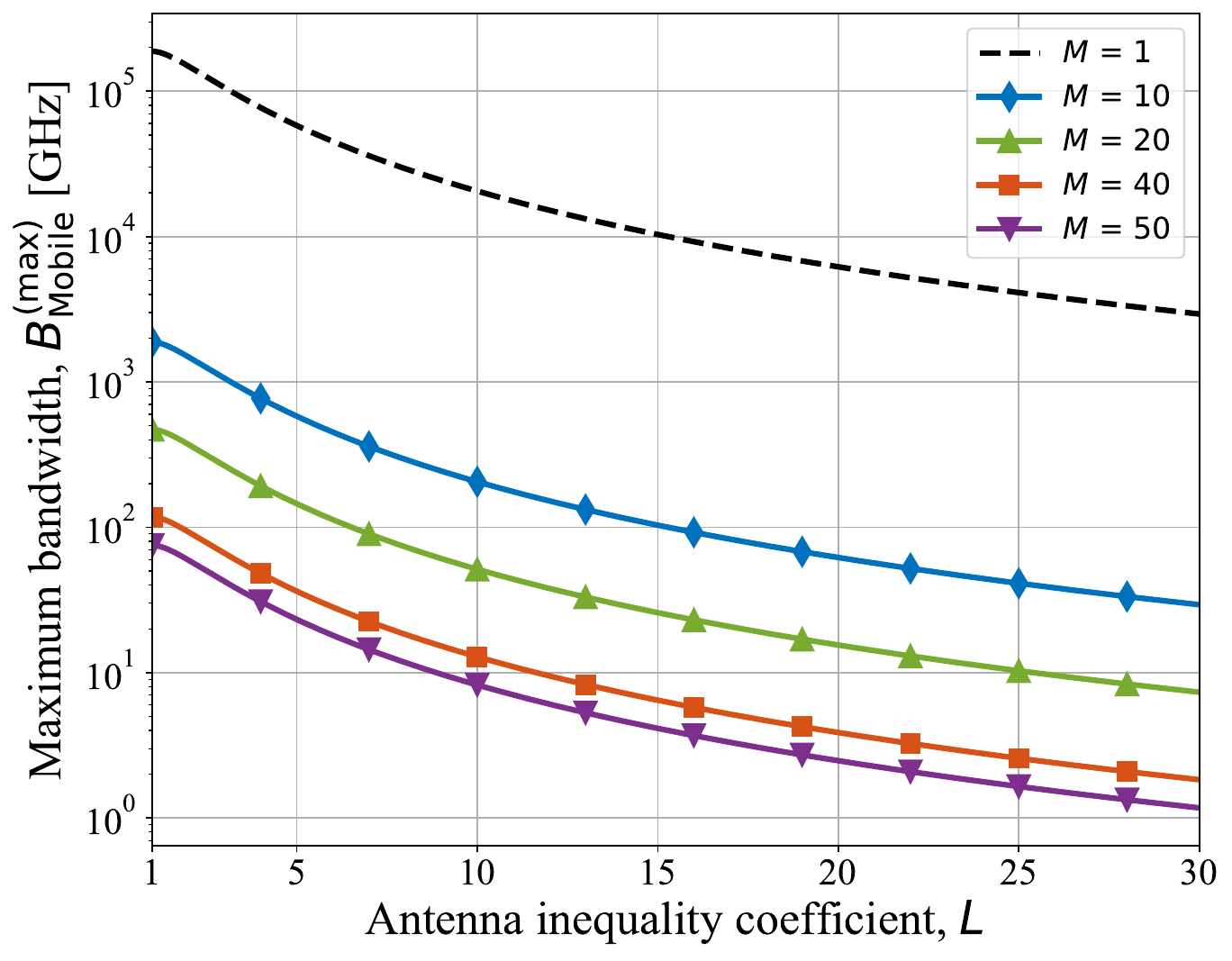}
  \label{fig:plot5}
 }
  \subfigure[Difference in antenna sizes at AP and UE]
 {
  \includegraphics[height=0.243\textwidth]{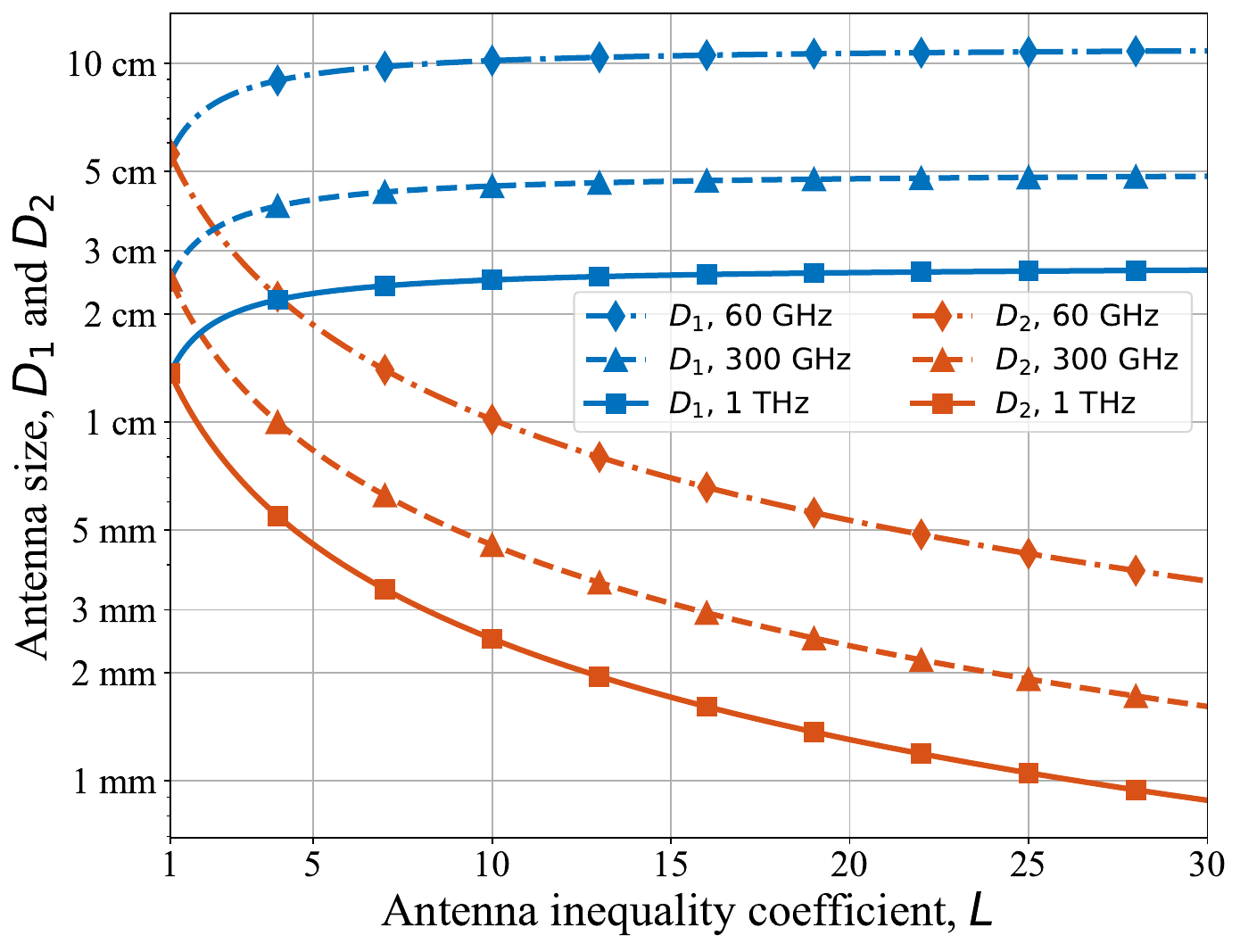}
  \label{fig:plot6}
 }
 \vspace{-2mm}
 \caption{The maximum achievable bandwidth for the far-field \emph{mobile} THz link and the corresponding antenna sizes.}
 \label{fig:mobile}
 \vspace{-4mm}
\end{figure*}

\newpage
\section{Numerical Results}
\vspace{-1mm}
\label{sec:numerical_results}
In this section, we numerically analyze the limits for stationary and mobile THz systems developed in Section~\ref{sec:analysis}.

\subsection{Stationary THz Wireless Link}
\label{sec:numerical_stationary}
\vspace{-1mm}

Figure~\ref{fig:stationary} presents the possible restrictions for a stationary THz link designed to operate exclusively in the far-field zone. Here, Tx and Rx have identical arrays of $N \times N$ elements and $D \times D$\,m$^2$ dimension. In Fig.~\ref{fig:plot1}, the maximum achievable bandwidth, $B^{\text{(max)}}_{\text{Stationary}}$, is a function of the transmit power, $P_{\text{Tx}}$ and the noise figure at the receiver, $N_{\text{F}}$\footnote{The maximum bandwidth here does NOT depend on the distance $d$ or the wavelength $\lambda$. Some values of $B^{\text{(max)}}_{\text{Stationary}}$ even exceed the THz frequencies, so the bandwidth is not limited at all. However, the effect of larger $d$ and $\lambda$ comes with larger arrays required (greater $D$ and $N$), see Fig.~\ref{fig:plot3}.}. The target SNR value at the receiver, $S_{\text{L}, \text{dB}}$ is set to $30$\,dB, sufficient to maintain a reliable high-rate backhaul link~\cite{priyangshu_thz_weather}. The dashed curve for $N_{\text{F}}=0$\,dB is a reference scenario with idealistic hardware.

We observe that greater available bandwidth can be achieved with both higher transmit power and better receiver hardware (lower $N_{\text{F}}$). Importantly, the absolute values are of \emph{hundreds of gigahertz and even several terahertz} for practical $P_{\text{Tx}} \geq 50$\,mW, already experimentally demonstrated at THz~\cite{sen2020teranova}. We observe a similar trend and even greater absolute values of $B^{\text{(max)}}_{\text{Stationary}}$ in Fig.~\ref{fig:plot2} when relaxing the SNR requirement to a lower value $S_{\text{L}, \text{dB}} < 30$\,dB. \emph{Hence, neither the transmit power nor the realistic noise figure or SNR requirement necessitates the \underline{stationary} system to work in the near field zone.}

Later, in Fig.~\ref{fig:plot3}, we explore the sizes of the antenna apertures $D$, at the Tx and the Rx, to achieve the maximum bandwidth while remaining in the far field. Here, the required antenna size decreases with the frequency, since smaller wavelengths increase the relative gain from a device by a factor of $(1/\lambda)^{2}$, see (\ref{eq:condition2}), while the near-field distance increases as a factor of $1/\lambda$, see (\ref{eq:condition1}). In contrast, while the communication range, $d$, does not impact the achievable bandwidth values in Fig.~\ref{fig:plot1}--\ref{fig:plot2}, greater $d$ demands larger arrays on both the Tx and the Rx to compensate for the spreading losses.

Nonetheless, observing the absolute values of $D$, we notice that for a typical $200$\,m--long backhaul link at $300$\,GHz, roughly a $10$\,cm $\times$ $10$\,cm antenna array is required, with the sizes further decreasing as the frequency keeps growing. The actual values of $D$ would be even smaller in practice, as the ones presented maintain the \emph{maximum bandwidth}, $B^{\text{(max)}}_{\text{Stationary}}$, which we have earlier observed to be several orders of magnitude higher than realistically required or utilizable. 

\textbf{Hence, we conclude that in most of the target practical scenarios, a \underline{stationary} THz wireless link can be optimized to work exclusively in the THz far field while still delivering substantial bandwidth using reasonably-sized antennas.}

\subsection{Non-Stationary THz Links in the Far Field}
\label{sec:numerical_mobile}
When switching from stationary to mobile THz communications, we must account for three key limitations.

First, the need to maintain a link over a range of communication distances $[d_{\text{min}}, d_{\text{max}}]$ instead of a single value of $d$. The SNR threshold must be satisfied at a larger $d_{\text{max}}$ value, while the antenna sizes of the mobile system are limited by the near-field border at a much smaller $d_{\text{min}}$. We utilize a mobility coefficient $M = d_{\text{max}} / d_{\text{min}}$ to account for this dynamic range. 
Second, a practical mobile system cannot assume the antenna size $D_{2}$ at a mobile node to be the same as the antenna size $D_{1}$, at a larger AP. We model this inequality through the antenna inequality coefficient $L = D_{1}/D_{2}$. Finally, the maximum uplink Tx power for a battery-powered mobile node is limited by a lower value than the one for a stationary backhaul link.

Figure~\ref{fig:plot4} presents the maximum achievable bandwidth of a far-field mobile THz system as a function of the mobility coefficient, $M$. The uplink transmit power of a mobile node is set to $23$\,dBm with the receiver noise figure at $10$\,dB. The practical range of $M$\footnote{A typical range of an indoor small cell varies from $0.5$\,m to $25$\,m ($M = 50$) and for an outdoor microcell -- in between $5$\,m and $200$\,m ($M = 40$).} is in the order of $40$ to $50$. In Fig.~\ref{fig:plot4}, we first observe that the maximum achievable bandwidth is expectedly decreasing with increasing $M$. Even for a practical value of $M = 50$, a non-stationary THz system can operate over \emph{few gigahertz} of bandwidth when using symmetric arrays at the UE and AP ($L = 1$), with high power front-ends. 

A similar observation is made from Fig.~\ref{fig:plot5} for a practical range of antenna inequality, $L$, in the order of $20$--$30$\footnote{The THz-AP antenna is in the order of up to dozens of cm$^2$, while the size of the THz-UE antenna is not exceeding a few mm$^{2}$.}.  As expected, the available bandwidth reduces as $L$ increases. However, if there is no mobility at all ($M=1$), \emph{several terahertz of bandwidth} can be occupied even when the inequality reaches $L=30$. 
It is worth noting that the corresponding array size at the access point $D_{1}$ and the UE $D_{2}$ ($d_{\text{min}} = 10$\,m) are still feasible, as shown in Fig.~\ref{fig:plot6} for different frequencies. 

\textbf{However, when \underline{both} $M$ and $L$ are set to their realistic values for a mobile THz communications scenario, the bandwidth becomes severely limited, as shown in Fig.~\ref{fig:plot4} and Fig.~\ref{fig:plot5} and further discussed in the next subsection.}

\subsection{Does 6G THz Wireless Have to Operate in Near Field?}
\begin{figure}[t!]
\centering
\includegraphics[width=0.9\columnwidth]{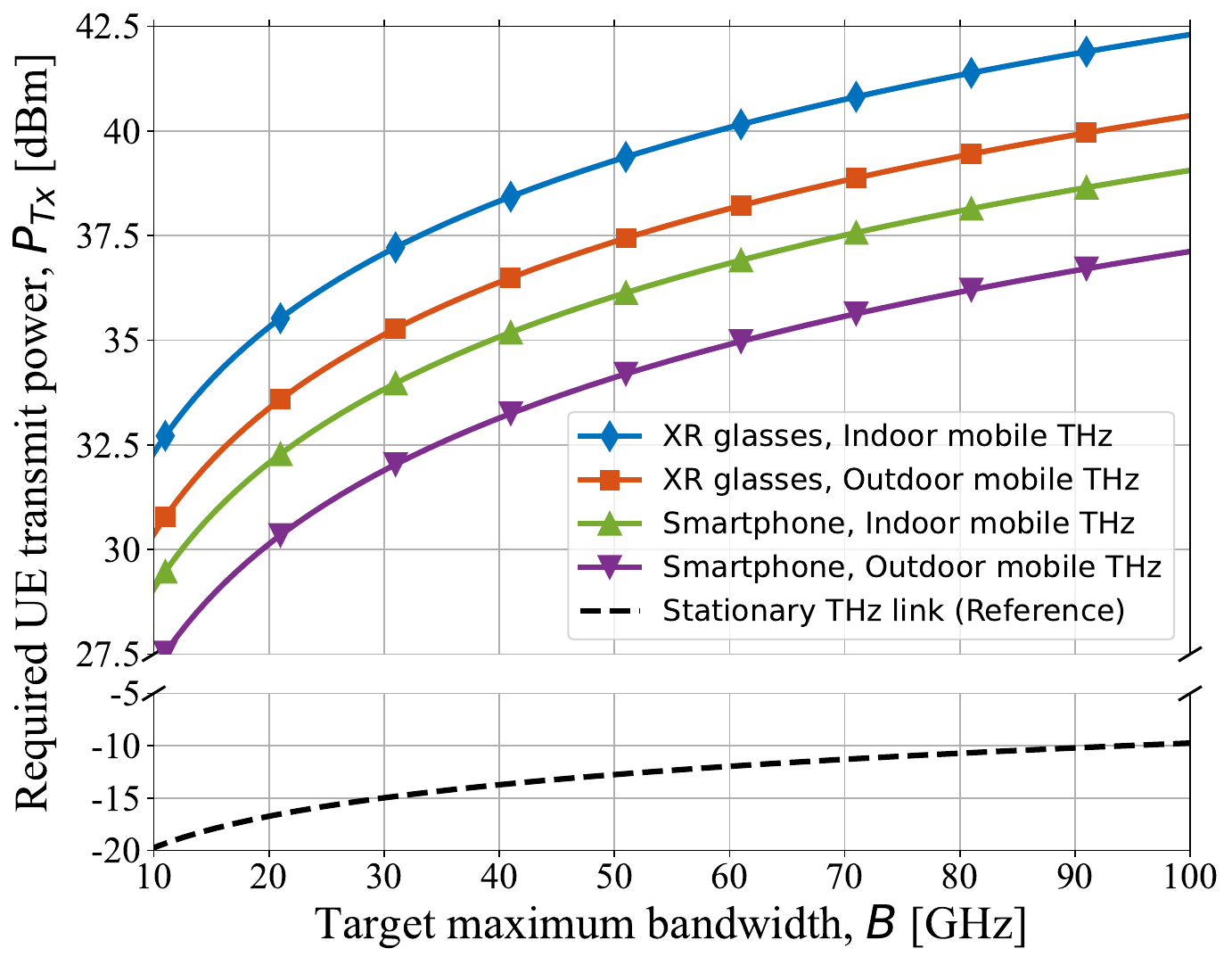}
\vspace{-2mm}
\caption{The maximum achievable bandwidth if the 6G radio is \underline{not} near-field. Contrasting \emph{stationary} 6G radio to \emph{mobile} 6G radio.}
\vspace{-5mm}
\label{fig:plot7}
\end{figure}

\vspace{-1mm}
In Fig.~\ref{fig:plot7}, we follow (\ref{eq:p_Tx_final_mobile}) and present the minimum required transmit power as a function of the desired bandwidth for stationary far-field THz wireless and mobile far-field THz wireless systems. The noise figure is $N_{\text{F}} = 10$\,dB and the target SNR, $S_{\text{L}, \text{dB}}$ is $20$\,dB\footnote{While the typical cell-edge SNR varies around $3$\,dB--$5$\,dB, this is accounting for NLoS propagation, fading, and other effects. Hence, the idealistic SNR value in ideal LoS should be at least $20$\,dB to keep a margin for NLoS, fading, weather, and other unfavorable propagation effects in practical deployments.}.
For mobile THz systems, we target two prospective form factors (smartphone, $L \approx20$, and XR glasses, $L\approx30$) in two realistic deployment scenarios (indoor access, $M\approx50$, and outdoor THz small cell, $M\approx40$).

We first observe from Fig.~\ref{fig:plot7} that a larger system bandwidth always requires greater transmit power. We also notice that for XR glasses, a small THz-UE form factor makes operating in the far field challenging (greater $P_{\text{Tx}}$ required), as a 
larger array at the THz-AP is required, increasing the near-field zone.

Most importantly, we study the absolute values of $P_{\text{Tx}}$ contrasting stationary and non-stationary THz links. 
Here, confirming the observations from Fig.~\ref{fig:stationary}, a stationary THz link can harness $100$\,GHz of bandwidth, operate in the far field, and with the transmit power of as low as just $-10$\,dBm. In contrast, non-stationary THz mobile links, especially involving wearable devices with physically small antennas, demand \underline{at least $30$\,dBm transmit power} to exploit \underline{$10$\,GHz of bandwidth} while avoiding near-field THz propagation effects. Such transmission powers are much higher than the state-of-the-art THz equipment can offer (even most stationary systems), would exceed present safety limits, and also notably challenge the energy efficiency of prospective mobile THz wireless access systems. De-facto, these are not feasible. Hence, the ``optimal'' sizes of the antenna arrays to keep the system exclusively far-field are not large enough. 

\textbf{Therefore, the bandwidth requirements dictate that prospective \underline{mobile} THz wireless communication systems \underline{will almost inevitably operate in the near field}, at least for portions of its coverage closer to the minimal distance.}

\section{Conclusions}
\label{sec:conclusions}
In this paper, we aim to clarify the need for near-field-specific (sub-)THz solutions in 6G-grade wireless networks. For this purpose, 
we develop a mathematical framework modeling an artificial idealistic case, where a stationary/mobile THz system aims to achieve the desired performance while working exclusively in the THz far field. We show that, importantly, such a system will be limited not by the central frequency (as per canonical near-field Fraunhofer distance~\cite{balanis2016antenna}), but by its maximum permissible bandwidth.

We then numerically elaborate the results illustrating that a stationary THz link can still operate over a notably wide bandwidth without dealing with near-field propagation. In contrast, the bandwidth limitations for realistic mobile THz links are considerably more stringent: as narrow as the bands already available in the 5G mmWave spectrum or even smaller. Hence, we conclude that while stationary THz links may operate in the near field if needed, but don't have to, mobile THz communication systems \emph{must} operate in the THz near field (at least for some distances) to be attractive for high-rate bandwidth-hungry use cases envisioned for 6G and beyond.

\appendices

\section{Near-field Distance Estimation for Two Square Antenna Arrays}
\label{app:fraunhofer_distance_array}

The conventional Fraunhofer distance is determined by~\cite{balanis2016antenna}:
\begin{align}
d \geq d_{\text{F}}\textrm{, where   } d_{\text{F}} = \frac{2 D^2}{\lambda},
\label{eq:app_a_d_f}
\end{align}
where $d_{\text{F}}$ is the length of the near-field zone, $D$ stands for the length of the largest physical dimension of the Tx antenna 
and $\lambda$ stands for the wavelength of the transmitted signal. This equation is applicable to a fixed-size Tx antenna and a point Rx (or, due to the reciprocity of the wireless channel, to the symmetric setup with a point Tx and a fixed-size Rx antenna). 

\begin{figure}[h!]
\centering
\vspace{-2mm}
\includegraphics[width=0.9\columnwidth]{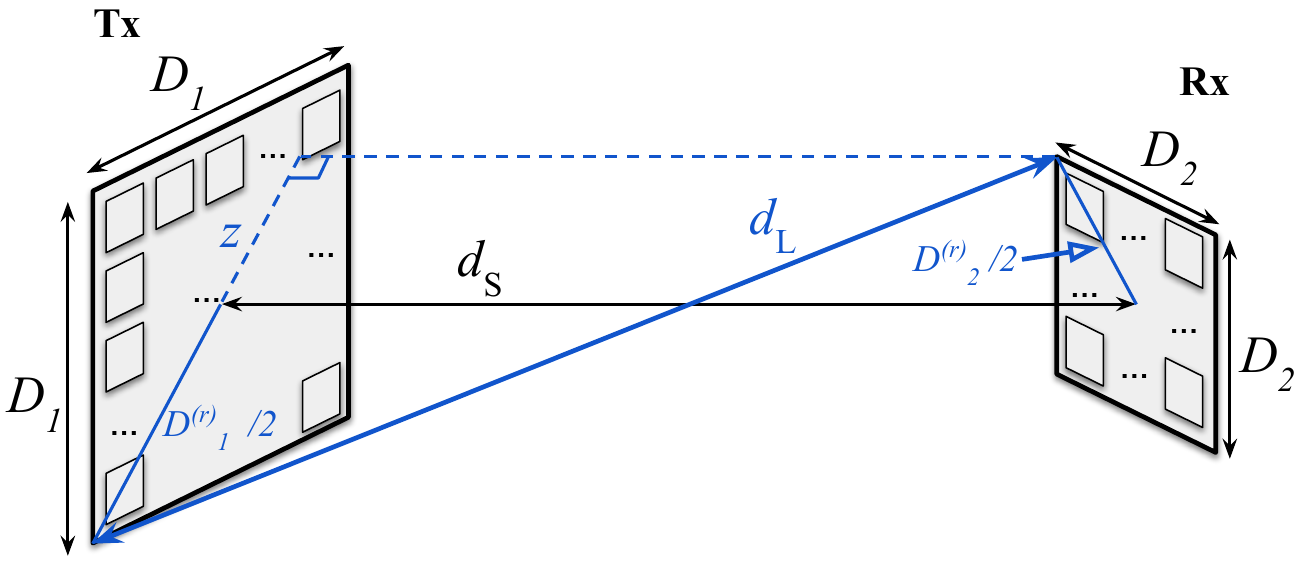}
\vspace{-2mm}
\caption{Calculating the analog for the Fraunhofer far-field distance in case of two 2D antenna arrays of sizes $D_{1} \times D_{1}$ and $D_{2} \times D_{2}$, respectively.}
\label{fig:app_a_1}
\vspace{-1mm}
\end{figure}

In this Appendix, we illustrate how (\ref{eq:app_a_d_f}) can be extended to the case of a Tx equipped with a planar antenna array of a physical $D_{1} \times D_{1}$ and an Rx equipped with a planar array of $D_{2} \times D_{2}$, as illustrated in Fig.~\ref{fig:app_a_1}.

In Fig.~\ref{fig:app_a_1}, we first identify the lengths of the shortest and the longest propagation paths ($d_{\text{S}}$ and $d_{\text{L}}$, respectively) as:
\begin{align}
d_{\text{S}} &= d_{\text{F}};\quad d_{\text{L}} = \sqrt{d^2 + z^2}, 
\label{eq:app_a_1}
\end{align}
where (from the setup geometry), $z = \frac{ D^{(r)}_{1} + D^{(r)}_{2} } {2} = \frac{D_{1} + D_{2} } {\sqrt{2}}$.

We then recall that, to meet the phase discrepancy limitation of $\pi/8$~\cite{balanis2016antenna}, the signal should propagate at maximum only $\pi/8$ longer in phase over $d_{\text{L}}$ than over $d_{\text{S}}$, giving:
\begin{align}
d_{\text{L}} = d_{\text{S}} + x,
\label{eq:app_a_2}
\end{align}
where $x = \frac{\lambda}{2\pi / \left(\frac{\pi}{8}\right)} = \frac{\lambda}{16}$ (the signal phase shift after traveling $\lambda$ meters is $2\pi$, so the signal phase shift is $\pi/8$ when propagating over $16$ times smaller distance).

Solving (\ref{eq:app_a_1}) and (\ref{eq:app_a_2}) together for $d_{\text{F}}$, we end up with:
\begin{align}
\frac{d_{\text{F}} \lambda}{8} + \frac{\lambda^2}{256} = \frac{ \left( D^{(r)}_{1} + D^{(r)}_{2} \right)^2} {4},
\label{eq:app_a_3}
\end{align}
where $\frac{\lambda^2}{256}$ is asymptotically smaller than other elements for high-frequency signals and can thus be omitted, leading to:
\begin{align}
d_{\text{F}} \approx \frac{ 2\left( D^{(r)}_{1} + D^{(r)}_{2} \right)^2} {\lambda} = \frac{ 4\left( D_{1} + D_{2} \right)^2} {\lambda}.
\label{eq:app_a_4}
\end{align}

It is important to note that, as per (\ref{eq:app_a_4}), the far-field distance is now determined by the physical dimensions of both the Tx and the Rx antenna arrays.
We utilize \eqref{eq:app_a_4} when determining the far-field distance in the main body of the paper.

\bibliographystyle{IEEEtran}
\bibliography{mobile_thz_math}

\end{document}